**Spin gaps in Transition Metal Dichalcogenide Nanoribbons with atomic Adsorbates**

Joshua O. Aggrey[1], Leonard Bleiziffer[2], Frank Hagelberg[3]*

Department of Chemistry, East Tennessee State University, Johnson City TN 37614 USA[1]

Institute of Theoretical Physics, Chinese Academy of Sciences, Beijing 100190, China[2]

Department of Physics, East Tennessee State University, Johnson City TN 37614 USA[3]

*Corresponding author: hagelber@etsu.edu

**ABSTRACT**

Edge-functionalized Transition Metal dichalcogenide nanoribbons of the zigzag type (zTMDCNRs) are explored in terms of their spin transmission properties. Specifically, systems of the type 5-zWXYNR+nA (X, Y = S, Se; n = 0 1, 2; A = H, B, C, N, O), involving five rows of a zWXY unit, are investigated as transmission elements between semi-infinite electrodes, to identify atomic adsorbates and adsorption conditions for maximizing the spin polarization of current traversing the ribbons. Janus counterparts of these units, asymmetric structures comprising a transition metal layer sandwiched by two different chalcogen layers, are included in this study. In all cases considered, density functional theory (DFT) modeling, involving the hybrid Heyd–Scuseria–Ernzerhof (HSE) exchange-correlation functional, is combined with the non-equilibrium Green's function (NEGF) approach to determine both spin and charge transport properties. The effect of the selected atomic absorbates on the geometric, electronic, and magnetic properties of 5-zWXYNR (X, Y = S, Se) is evaluated. A protocol to assess the spin-filtering capacity of 5-zWXYNR+n*ad* as a function of the nature and the density of atomic adsorbates, is formulated in terms of band structure analysis of the respective electrode units. Spin gaps emerging close to the Fermi energy of the electrode are shown to provide an effective predictor for the degree of current spin polarization achieved by any of the transmission systems studied here. For any adsorbate configuration considered, ferromagnetic (FM) as well as antiferromagnetic (AFM) ordering is examined, and the impact of the magnetic phase on the spin transport properties is discussed. A spin-selective negative differential resistance effect is identified for specific nanoribbon systems.

## 1. INTRODUCTION

Spintronics at the nanoscale, or nanospintronics [1], is a rapidly advancing branch of present-day nanoscience and technology. An important component of this initiative is the definition, characterization, and testing of novel nanostructures that might serve as elements of spintronics networks. In this context, research has focused on planar (2D), linear (1D), or point-like (0D) nanomaterials, where the dimensionality is given by the maximum number of periodic coordinates of a nanostructure.

As the general criterion, elements of spintronics circuits must support spin-dependent interactions and, more specifically, display potential to act as spin filters, spin valves, spin diodes, spin field-effect transistors, or spin-orbit torque devices [2-6]. In this work, we focus on the spin filtering efficiency of 1D transmission elements (nanoribbons), i.e., their capacity to generate high degrees of spin current polarization, which, in turn, gives rise to strong magnetoresistance effects, and thus, in marked spin valve behavior. Spin filtering activity is determined by *spin gaps* in the spintronic carrier material, i.e., the prevalence of electron bands with a unique spin orientation in the vicinity of the Fermi energy, $E_F$. In this context, a wide range of magnetic nanomaterials is currently considered as candidates for use as spintronic elements. These include InAs quantum dots [7] or nanowires [8], carbon nanotubes [9], monolayers like graphene or graphene-analogous systems [10, 11], or surfaces of topological insulators, such as $Bi_2Se_3$ [12]. In this work, we consider transition metal dichalcogenides (TMDCs) with TM = W as possible spin filtering or spin valve nanodevices.

TMDCs form stacked semiconducting compounds, consisting of TM sheets (TM = Ti, V, Cr, Tc, Hf, Ta, W, Mn, Zr, Nb, Mo, and Re) sandwiched by two chalcogen (X) sheets, where X = S, Se, Te [13]. Like graphene, TMDC monolayers may be fabricated by exfoliation. As typical two-dimensional (2D) materials, both share a high surface-to-volume ratio, which is of high interest for applications in the fields of catalysis or sensing [14]. Both materials display extraordinary mechanical strength and flexibility [15], and also rank high in terms of charge carrier mobility. While graphene conducts electrons almost without resistance at room temperature, TMDC monolayers exhibit lower electron mobility, which is, however, significantly higher than that of traditional semiconductors [16].

Another feature shared by graphene and TMDC monolayers is the absence of magnetism in their respective ground states at equilibrium, where TMDCs based on TM = Cr, V, possibly form exceptions from this rule [13]. Thus, to exploit the favorable materials properties of graphene and TMDC sheets for spintronics, one needs to induce magnetism in the respective pristine systems. This can be achieved by importing magnetic components – e.g., adatoms or substitutional impurities with finite magnetic moments – into the material of interest [17]. Another pathway is structural modification. Thus, geometric defects, such as vacant sites, may be created [18], or the dimensionality of the structure may be reduced from 2D to 1D [19]. Leveraging the second strategy, one may generate graphene nanoribbons (GNRs) or nanotubes (GNTs) from graphene or TMDC nanoribbons from TMDC monolayers. In the present work, the latter are examined as potential elements of nanospintronics.

Various computational studies suggest ground-state magnetism for nanoribbons derived from the TMDC phase of maximum stability (2H) [20-21]. Nanoribbons of both the armchair (aTMDCNRs) and the zigzag (zTMDCNRs) variety, corresponding to different geometric patterns of the ribbon edges, as illustrated in Figure 1, have been explored [22, 23]. From these investigations, the zigzag structure tends to be more stable than the armchair structure. Further, zTMDCNR systems turned out to exhibit ground-state magnetism, where magnetic phases are typically separated from the nonmagnetic phase by distinct margins of $\Delta E \geq 0.1$ eV [19]. Along zigzag edges, ferromagnetic (FM) coordination has been established [19, 21], while ferromagnetic or antiferromagnetic (AFM) configurations have been found to prevail between the top and the bottom edge of zTMDCNRs, depending on the specifics of the TM and DC combinations considered [24].

*Figure 1*. Geometrical structure of tungsten disulfide nanoribbons ($WS_2NRs$). (a) Crystal geometry of tungsten disulfide, $WS_2$; (b) Armchair nanoribbon ($aWS_2NR$) based on a $WS_2$ monolayer; (c) Zigzag nanoribbon ($zWS_2NR$) based on a $WS_2$ monolayer. Shown is the five-layered $zWS_2NR$ ($5-zWS_2NR$) species analyzed in this work.

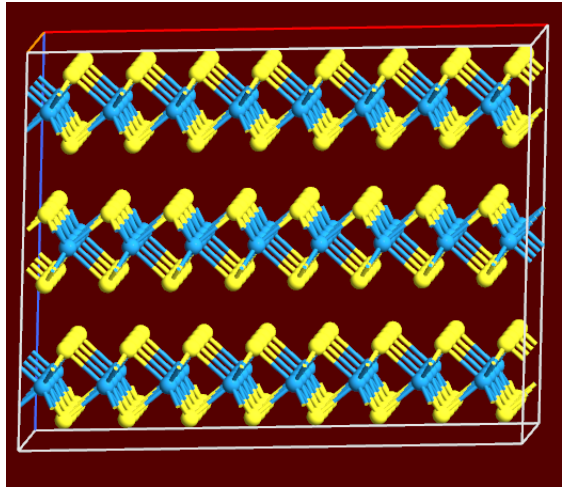

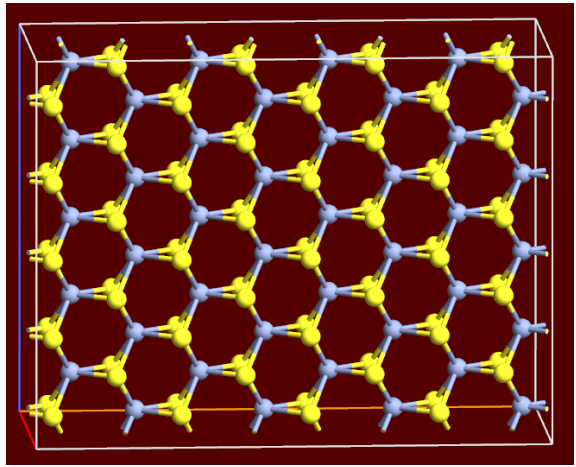

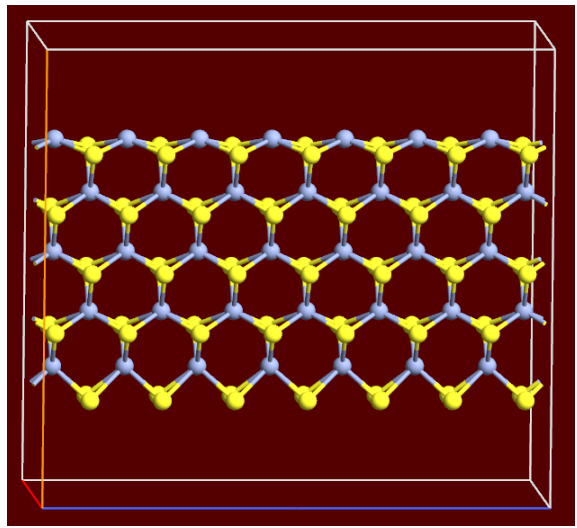

Various modes of manipulating the magnetic properties of zTMDCNRs have been described. Thus, applying longitudinal strain on the ribbon has been established as a way to convert the two competing magnetic phases, FM and AFM, into each other [24]. Further, Janus structures, comprising a TM layer sandwiched by two layers with differing DC atoms [24-27], such as S and Se, have been analyzed. The associated breaking of the TMDC mirror symmetry naturally induces an electric field perpendicular to the TM sheet, leading to numerous distinctive physical phenomena of impact on the magnetism of the nanoribbon, including increased interlayer coupling, Rashba spin splitting, and valley polarization [28]. Laterally or vertically applied gate fields may be employed to modify this effect [24].

In this contribution, we explore the potential of edge functionalization with atomic adsorbates to alter the spintronics properties of zTMDCNRs. This inquiry is motivated by the high reactivity of the TM zigzag edge, exceeding that of the chalcogen-terminated edge [29, 30]. Thus, in terms of practical applications, edge-saturated zTMDCNRs are of higher relevance than the pure units. The magnetic features are expected to change in response to TM functionalization, as the Bloch states of the weakly bound zTMDCNR bands are largely composed of incompletely saturated d-orbitals of the TM species [24, 30] and thus sensitively affected by chemical edge modification. Here, we (1) examine the impact of light, non-metallic atomic adsorbates on the magnetism, and specifically the spintronic properties, of WXY nanoribbons, and (2) outline protocols for adsorbate addition, involving the nature, the concentration, and the adsorption geometry of the considered adatoms, for maximizing the spintronic effect.

We choose WXY (X, Y = S, Se) as a TMDC reference system and investigate nanoribbons of the form 5-zWXYNR+n*ad* (X, Y = S, Se; n = 0, 1, 2; A = H, B, C, N, O), composed of five vertically stacked zWXY units. Systems of interest are identified by inspecting the geometric and electronic structures of 5-zWXYNR+n*ad* electrode cells and analyzing them with respect to their spin transport properties. To this end, we compute their spin-resolved transmission spectra. In a third step, the electrode cells are expanded into spintronics nanodevices. The current-voltage profiles of these materials are computed, yielding direct information about the degree of current spin polarization in the considered voltage interval. From these results, conclusions are drawn with respect to the spin filtering activity and magnetoresistance behavior of the analyzed units [31].

## 2. METHODS

All nanoribbons included in this work were analyzed in terms of their electronic structure and their transport properties. In what follows, we provide a summary of procedures that were applied in both types of investigation.

### 2.1. Electronic Structure.

The electronic system was modeled by density functional theory (DFT) computation in conjunction with a localized basis within periodic boundary conditions. PseudoDojo ONCV pseudopotentials, i.e., pseudopotentials of scalar-relativistic, optimized norm-conserving Vanderbilt type, with associated localized numerical basis sets [32], were adopted to make the calculations feasible. Recent research [33] demonstrated the good reproducibility of PseudoDojo-generated results on heterostructures involving TMDCs across a variety of pseudopotential choices, among them the Projector Augmented Wave (PAW) [34] approach.

Only electrons beyond the nearest rare gas configuration were included in the DFT treatment, reducing the W shell to 14 and the chalcogen shells to 6 (S) and 16 (Se) electrons. Specifically, the W shell was represented by the electronic configuration [Xe+4f14] 5s2, 5p6, 5d4, 6s2, and the Se shell by [Ar] 3d10, 4s2, 4p4, where the labels following the square brackets indicate the electrons included in the DFT treatment. Numerical atom-centered basis sets [35] at the Double Zeta Polarized (DZP) level were employed. A test computation for a selected system demonstrated good agreement with results obtained by use of a more complex basis, involving the Triple Zeta Polarized (TZP) level (see section *S-A* of the Supporting Information file).

The electronic structure calculations were performed by use of spin-polarized DFT, involving two functionals, namely the Perdew−Burke−Ernzerhof (PBE) exchange-correlation functional [35] for geometry optimizations, and the Heyd-Scuseria-Ernzerhof (HSE) functional [36] for band structure and transmission (see 2.2) computations. As the crucial quantities of this study, i.e., spin-resolved transmission spectra and current spin polarizations (see II.1), are crucially dependent on an accurate representation of the band structure regime close to the Fermi energy, the PBE functional – used widely in electronic transport calculations – is insufficient for yielding quantitative results in the present context. Thus, the PBE functional is known to severely underestimate the band gaps of semiconductors and insulators [37]. This deficiency is attributed to the (semi-)local exchange inherent in the PBE approach and may be alleviated by use of the HSE functional that corrects for self-interaction errors and tends to predict band gaps closer to experiment, typically within 0.2–0.4 eV [38].

We used this approach to represent devices of the form zWXYNR+n*ad* (X, Y = S, Se, *ad* = adatom, n = 1, 2 – see Figure 2 (e-f)), consisting of a finite transmission element wedged between semi-infinite electrodes. The latter were designed as junctions of two zWXYNR+n*ad* unit cells (Figure 1 (a-d)). Geometry optimization within periodic boundary conditions was performed to define electrode equilibrium structures. Forces on the nuclei (electronic energies) were constrained to converge with an accuracy of 0.05 eV/Å ($1.0 \times 10^{-6}$ eV). The unit cell lattice parameter in the transport direction was included in the geometry optimization procedure. The density mesh cutoff energy was chosen as 110 Hartree, and a wave number mesh of dimension $1 \times 1 \times 150$ was applied, where the last factor denotes the wave number subdivision in the transport direction.

Device equilibrium structures, as shown in Figure 2(e-f), were found by subjecting the transmission elements to geometry optimization, except for two segments adjacent to the electrodes. These were constrained to repeat the structure of the electrode unit cell. Adjacent images were separated by vacuum layers, chosen sufficiently spacious to keep the interaction between them negligible.

**2.2. Transport Calculations**.

Electron transport was simulated in the framework of the nonequilibrium Green's function (NEGF) theory, as implemented in the code QuantumATK [39-41]. Specifically, we modeled the spin-dependent current through the device in the framework of the Landauer-Büttiker approximation [42] according to:

$$I_\sigma(V_{bias}) = \frac{e}{h} \int T_\sigma(E, V_{bias})[f_L(E - \mu_L) - f_R(E - \mu_R)] \, dE \qquad (1)$$

Here, the expressions $f_i (E - \mu_n)$ (n = L, R) refer to the Fermi−Dirac distribution functions for the left and the right electrode, respectively, which depend on the electrochemical potentials $\mu_n$ (n = L, R). These, in turn, define the bias voltage across the transmission element, $V_{bias}$:

$$\mu_{L/R}(V_{bias}) = \mu_n(0) \pm \frac{eV_{bias}}{2} \qquad (2)$$

The symbol $T_\sigma$ represents the spin-resolved transmission, a function of the energy and the bias:

$$T_\sigma(E, V_{bias}) = \text{Tr}\{\Gamma_L G \Gamma_R G^\dagger\}. \qquad (3)$$

The left-hand side of this equation is the trace over the matrix $\Gamma_L G \Gamma_R G^\dagger = t t^\dagger$, where t stands for the transmission matrix whose elements $t_{nm}$ denote the amplitudes for the transition of an electron from the state n of one of the two electrodes into the state m of the other [43]. The symbols $\Gamma_i$ (i = L, R) are defined as

$$\Gamma_n = i(\Sigma_n - \Sigma_n^\dagger) \qquad n = L, R, \qquad (4)$$

i.e., as the anti-Hermitian components of the self-energy $\Sigma_n$ for the left (n=L) and the right (n=R) contact. The quantity G in relation (3) refers to the energy-dependent matrix of the Green's function for the transmission element. The approximation (1) is based on the ballistic assumption [44], which is justified in the present case, since the length of the transmission element is in the nanometer regime [45]. The effect of a finite bias on the electrostatic potential within the transmission element was included by incorporating the Poisson equation into the DFT treatment [46, 47].

## 3. RESULTS AND DISCUSSION

In the following, we will first introduce the electrode units cell adopted for systems of the form zWXYNR+n*ad* (X, Y = S, Se, n = 0, 1, 2, *ad* = H, B, C, N, O )  (3.1) and then turn to the nanodevices based on them (3.2). After summarizing the geometric structures of the electrodes (3.1.1), we focus on the electronic properties of zWXYNR+n*ad* (n = 0, 1) in terms of band structure profiles (3.1.2A(i)). Based on these findings, we consider the spin gap formation mechanisms (3.1.2A(ii)) and the transmission spectra (3.1.2A(iii)) characteristic of these units. Along similar lines, we present our results for zWXYNR+2A (3.1.2B). In subsection (3.2), we draw conclusions with respect to the current spin polarization achievable by zWXYNR+nA devices.

### 3.1 Electrodes

#### 3.1.1 Geometric structure

Four geometric prototypes were included, differing from each other by their edge adsorption patterns. Specifically, we distinguish between electrodes with one or two adatoms in an on-top or bridge position, located above the midpoint between the two W atoms in the upper edge of the unit cell. These geometric classes are presented in Figure 2 for the special case of H adatoms attached to 5-WS2NR edges. All of them are based on a two-cell electrode model.

*Figure 2*: The six geometric prototypes of zWXYNR+n*ad* (X, Y = S, Se, A = adatom, n = 1,2) considered here, by the example of X = Y = S, and *ad* = H. (a) – (d): the four electrode edge adsorption patterns selected for comparison. The panels show (a) a single adatom in an on-top position above a W atom (5-WS2NR+$H_{top}$), (b) two adatoms in on-top positions (5-WS2NR+$2H_{top}$), (c) a single adatom in bridge position between the two W atoms (5-WS2NR+$H_{bridge}$), and (d) two adatoms in bridge positions (5-WS2NR+$2H_{bridge}$). (e), (f): the device geometries based on the two bridge configurations, i.e., the single adsorbate (5-WS2NR+H, (e)) versus the double-adsorbate structure.

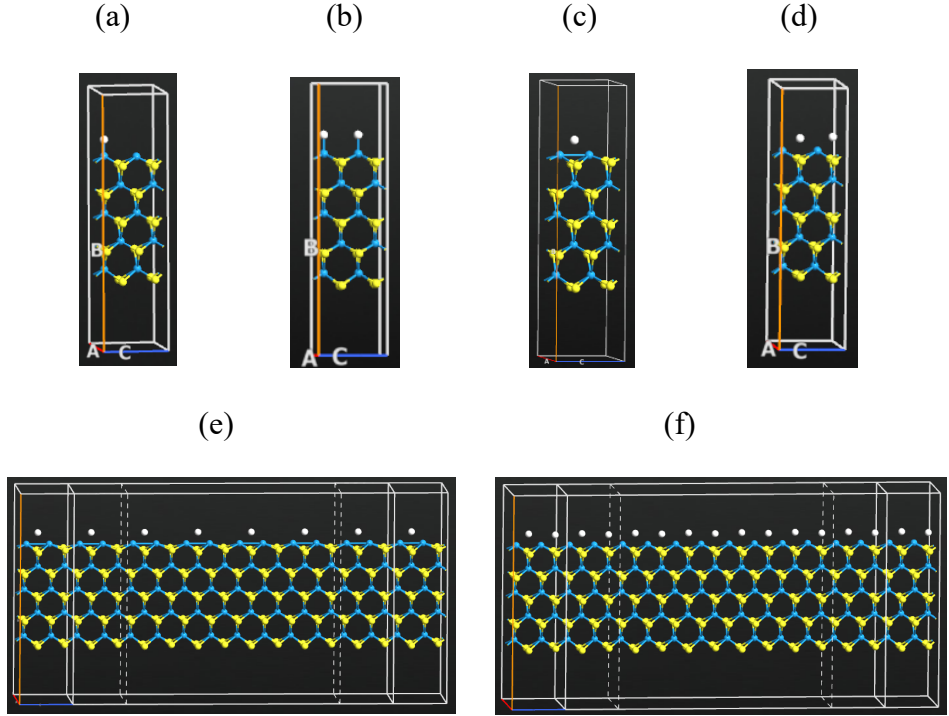

(a) (b) (c) (d)

(e) (f)

We compared these four alternatives in terms of their stabilities, i.e., their total energies at ground state equilibrium. The results are summarized in Table 1.

*Table 1*: Total energies of the four prototypes included in Figure 2 at optimization minima. The structure labels are as in Figure 2.

| Structure | Total Energy [eV] |
| --- | --- |
| 5-WS2NR+$H_{top}$ | -26014.2069 |
| 5-WS2NR+$H_{bridge}$ | - 26015.0679 |
| 5-WS2NR+2$H_{top}$ | -26031.3538 |
| 5-WS2NR+2$H_{bridge}$ | - 26031.5805 |

For both single and double adsorption, the adatom bridge positions turn out to yield lower total energies than adatom on-top positions. In view of this finding, we limited our electrode analysis to adatom bridge positions. When referring to these units, we will drop the subscript bridge and denote them, for simplicity, with zWXYNR+n*ad* (X, Y = S, Se; n = 1, 2; *ad* = H, B, C, N, O). In the following section, we assess electrode systems of this type with respect to their potential interest for applications in spintronics. This requires comparing them in terms of their magnetic properties.

### 3.1.2 Electronic structure

The initial step of our search for 5-zWXYNR+n*ad* (X, Y = S, Se; n = 0-2; *ad* = H, B, C, N, O) devices of relevance for nanospintronics involves analyzing 5-zWXYNR electrodes within periodic boundary conditions. Of specific low-energy bands of 5-zWXYNR electrodes are expected to provide valuable predictors for the spin transmission capabilities of the respective 5-zWXYNR devices.

The related research is guided by the following questions:

(a) Do the band structures of pristine 5-zWXYNR electrodes exhibit *spin gaps* in the Fermi energy regime, i.e., bands with distinct spin polarization?

(b) What variations in pristine 5-zWXYNR electrodes, such as changes of their chemical composition or admission of defects, give rise to spin gap formation?

Addressing these questions, we will consider first electrodes of the type 5-zWXYNR+ad, involving a single adsorbate atom at the upper edge of the structure (3.12A), and then turn to the double-adsorbate case (3.12B)

(A) Electrodes of the form zWXYNR+n*ad* (X, Y = S, Se; n = 0, 1; *ad* = H, B,C,N,O)

(i) Spin gaps in band structure profiles

Focusing on the 5-WS2NR electrode, and comparing the total energy values for the two alternative magnetic configurations, AFM and FM, we find a slight energetic advantage, in the meV regime, for the former (see Table 2). This is in keeping with previous assessments of zTMDCNR magnetism [19]. Figure 3(a, b) shows the band structure of the AFM ground state of 5-zWS2NR in an interval of [-1.0, +1.0] eV around the Fermi energy, $E_F$. The results in Figure 3(a) (Figure 3(b)) were evaluated by use of the SGGA (HSE) functional. The two procedures show very substantial deviations with respect to the band structure in the vicinity of $E_F$. In particular, the HSE06 functional yields a marked expansion of the bands close to $E_F$, with a direct energy gap of $\Delta E = \sim 0.25$ eV, when compared with the SGGA treatment. This observation agrees with the consensus that SGGA tends to underestimate energy differences between valence and conduction bands [48] while HSE06 offers a correction of this shortcoming [32, 35]. The following discussion will thus be based on HSE06 rather than SGGA computations.

*Figure 3*: Band structures of 5-zWXYNR+*ad* (X, Y = S, Se, *ad* = H, B, C, N, O) electrode systems in the energy range [-1.0, 1.0] eV. Shown are: (a) zW2NR (AFM, SGGA calculation), (b) zW2NR (AFM, HSE06 calculation), (c) zW2NR+H (AFM), (d) zW2NR+B (FM), (e) zW2NR+C, (f) zW2NR+N (g) zW2NR+O (h) WSeS (i) WSeS+H (AFM), (j) WSeS+C (FM). In each case, the magnetic configuration of lowest energy is selected. All structures, except case (a), were computed using the HSE06 functional. Bands with spin alpha (beta) polarization are shown in red (blue).

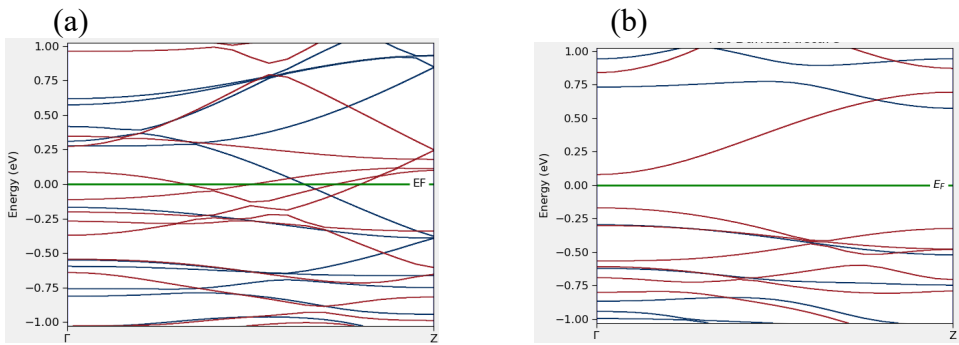

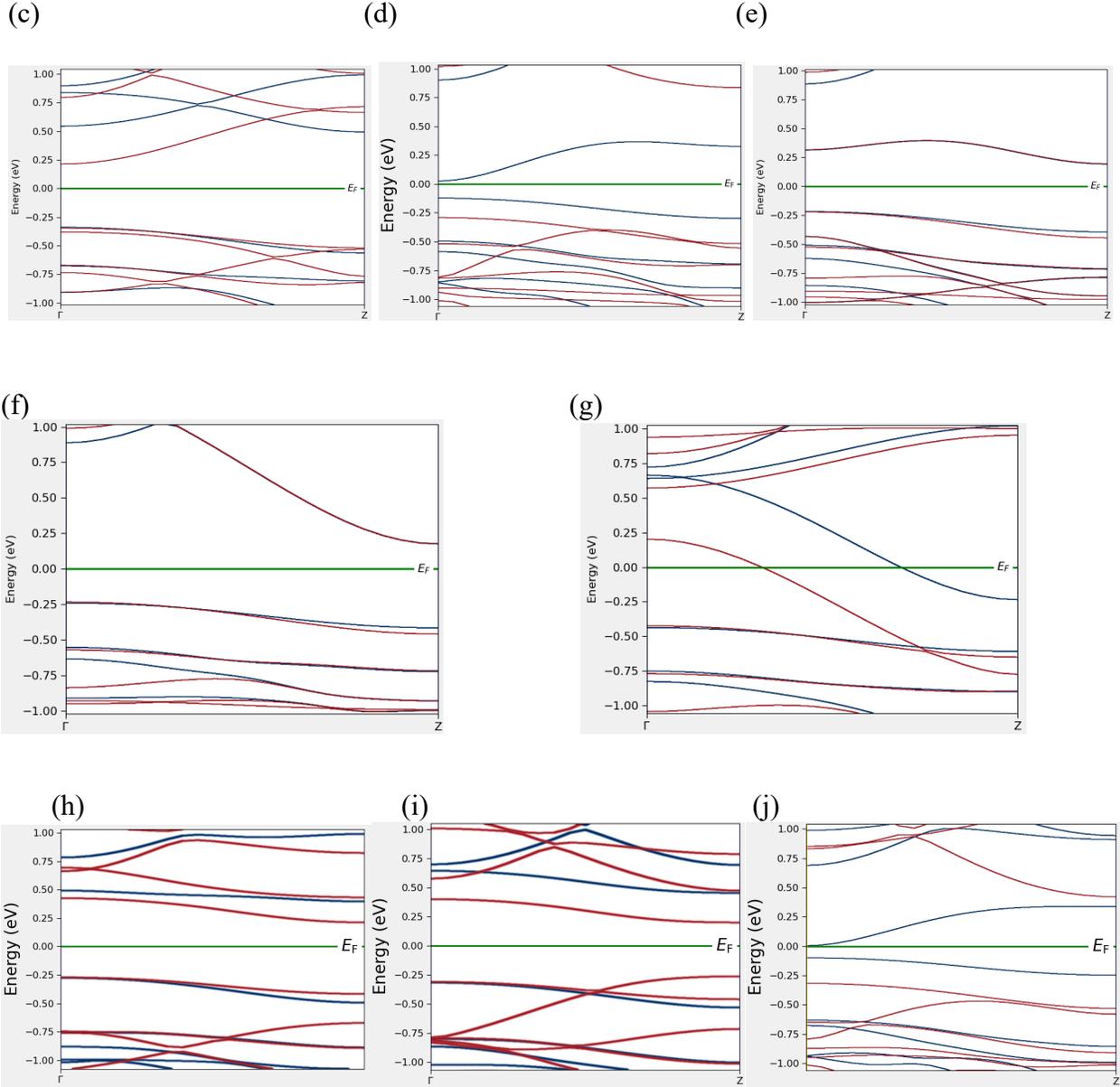

*Table* 2: Energetic and magnetic parameters of double-cell 5-zWXYNR+*ad* (n = 0, 1; X, Y = S, Se; A = H, B, C, N, O). The labels AFM (antiferromagnetic) and FM (ferromagnetic) refer to settings chosen for the initial spin conditions. For each system, the magnetic phase of higher stability is listed first.

| System | Magnetic Phase | E(HSE)[eV] | Net spin population on ad | Energy gap [eV] |
|---|---|---|---|---|
| 5-WS$_2$NR | AFM | -25997.6484 | | 0.25 |
| | FM | -25997.6480 | | 0.24 |
| 5-WS$_2$NR+H | AFM | -26015.0681 | 2.50 10$^{-3}$ | 0.55 |
| | FM | -26015.0679 | 1.45 10$^{-3}$ | 0.56 |
| 5-WS$_2$NR+B | AFM | -26076.4211 | -4.00 10$^{-4}$ | 0.53 |
| | FM | -26076.4211 | -1.50 10$^{-4}$ | 0.52 |
| 5-WS$_2$NR+C | FM | -26159.0627 | -4.70 10$^{-3}$ | 0.15 |
| | AFM | -26159.0610 | -3.72 10$^{-3}$ | 0.15 |
| 5-WS$_2$NR+N | FM | -26278.7608 | <1.00 10$^{-3}$ | 0.43 |
| | AFM | -26278.7603 | 1.00 10$^{-3}$ | 0.43 |
| 5-WS$_2$NR+O | AFM | -26446.9123 | -6.21 10$^{-3}$ | 0 |
| | FM | -26446.9123 | -6.02 10$^{-3}$ | 0 |
| 5-WSeSNR | AFM | -55123.6617 | | 0.63 |
| | FM | -55123.6616 | | 0.63 |
| 5-WSeSNR+H | AFM | -55138.7873 | -2.00 10$^{-2}$ | 0.46 |
| | FM | -55138.7821 | -2.00 10$^{-2}$ | 0.50 |
| 5-WSeSNR+C | FM | -55282.6896 | -1.90 10$^{-2}$ | 0.10 |
| | AFM | -55282.6892 | -1.70 10$^{-2}$ | 0.10 |

From our analysis, the band structure of 5-WS$_2$NR is characterized by an extended spin gap in the vicinity of E$_F$, encompassing an interval of 0.86 eV. This finding raises the expectation that this system, when developed into a spintronic device, might yield currents with high degrees of spin polarization. At the same time, the Fermi energy is located within an energy gap of 0.25 eV, which limits the size of the current delivered by pure 5-zWS$_2$NRs in response to small voltages, i.e., smaller than 0.25 V. The appearance of a small energy gap in the 5-zWS$_2$NR band structure may be rationalized by Bloch function analysis of the two frontier bands, i.e., the highest-lying valence band (HVB) and the lowest-lying conduction band (LCB), which represent a bonding and an antibonding arrangement of two neighboring W atoms at the ribbon top (see section *S-b* of the Supporting Information file). A metallic solution for 5-WS$_2$NR was identified as well. (see section *S-C* of the Supporting Information file). It was found to be higher in total energy than the semiconducting solution by a 67.2 meV.

Exploring possible electrode modifications that might likewise display a distinct spin gap but a lower, or even vanishing, energy gap, we considered the structural type shown in Figure 2(b), involving adsorbate atoms *ad* (*ad* = H, B, C, N, O) in a bridge position between the two W locations (see Section 3.A.1). The adsorption equilibrium geometries for 5-WS$_2$NR+*ad* (*ad* = H, B, C, N, O) are shown in Figure 5, along with the respective spin densities. As we attach an H adsorbate atom in the bridge position, the direct energy gap of the pure 5-WS$_2$NR system increases by more than a factor 2, widening to $\Delta E = 0.55$ eV (see panel 2(c)). Above the gap, one notices a 0.25 eV wide zone where alpha spin orientation, associated with the lowest conduction band, prevails. Exchanging the H for a B adatom, one retains a substantial energy gap but removes the spin gap, as the bands closest to $E_F$ display almost no net spin polarization (panel 3(d)).

For *ad* = B, we find a crossover of the most stable state from AFM to FM. However, for all units of the form 5-WS2NR+n*ad* (n = 0, 1, *ad* = H, B), the energy difference between the two magnetic alternatives turns out to be minute ($\Delta E$ in or below the meV range). In terms of energy bands, the difference between them is, in most cases, a spin orientation switch for one band. While this transition does not give rise to a marked change in total energy, it can have substantial consequences for the spintronic features of the ribbon, as may be demonstrated by the system with *ad* = C.

The band structure profile of 5-WS$_2$NR+C is indicative of a ribbon that holds major interest for spintronics applications. A marked spin gap, enclosing the Fermi energy, and extending over an interval of 0.6 eV (panel 2(e)), encloses here a narrow energy gap of 0.15 eV.

The asymmetric placement of the HVB and the LCB with respect to the Fermi energy in both magnetic states suggests that 5-WS2NR+C may yield substantial spin polarization at high current magnitudes. Thus, the respective energies with respect to the Fermi energy are 2.61 10$^{-2}$ eV (LCB) versus -1.22 10$^{-1}$ eV (HVB). The LCB is therefore expected to prevail strongly over the HVB in defining the spin polarization degree of current traversing the ribbon at low voltage.

As a caveat, assessing this unit in terms of its potential use as a spintronics element must include not only the magnetic configuration of lowest energy (FM, see Table 2) but also the near-degenerate AFM alternative. The band structures for both magnetic versions are shown in Figure 4 for 5-zWS$_2$NR+C, 5-zWSeSNR+C, and 5-zWSeSNR+B. From this comparison, the spin label of the HVB changes from beta(↓) to alpha(↑), since the HVB↑ and HVB↓ bands exchange their spin orientations as one goes from FM to AFM order. For 5-zWSeSNR+B, these two bands are near-degenerate, and so, the effect of this rearrangement is minimal. For the systems with a C adsorbate, however, the two bands are distinctly separated, and the impact on the magnetic structure is more distinct in these cases. For 5-zWS$_2$NR+C, the ground state energy difference between the two magnetic states amounts to only 2.0 meV. Thus, both will occur with almost equal populations. When deriving the current spin polarization for 5-WS$_2$NR+C devices, one therefore has to take the thermally weighted average between the results for FM and AFM order, i.e., between a system with an extended spin gap around E$_F$ (FM order) and one with two spin gaps of opposite spin orientations for E < E$_F$ and E > E$_F$. As the thermal weights for both states are approximately equal, the contribution of the latter system is expected to compromise the overall current spin polarization. The alternative of exposing the ribbon to an external magnetic field strong enough to lift the energy difference between the competing states beyond the thermal energy would, at room temperature, require a magnetic field in the order of 500 T, which could not be neglected in any modeling attempt and would be forbidding for most applications.

*Figure 4:* Band structures of 5-zWXYNR+*ad* (X, Y = S, Se, *ad* = B, C) electrode systems in the energy range [-1.0, 1.0] eV. The left (right) panels show the band structure profiles under FM (AFM) conditions. Included are 5-zWS$_2$NR+B (a), 5-zWS$_2$NR+C (b), and 5-zWSeSNR+C (c).

(a)

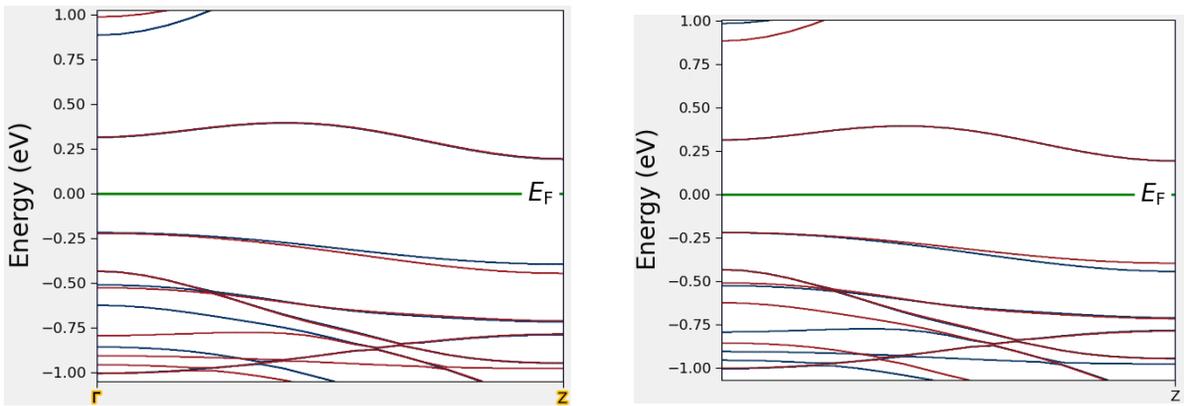

(b)

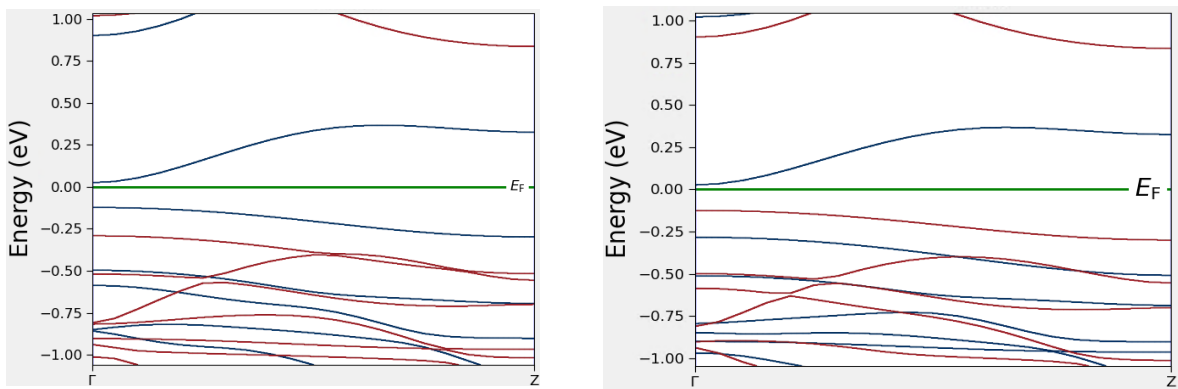

(c)

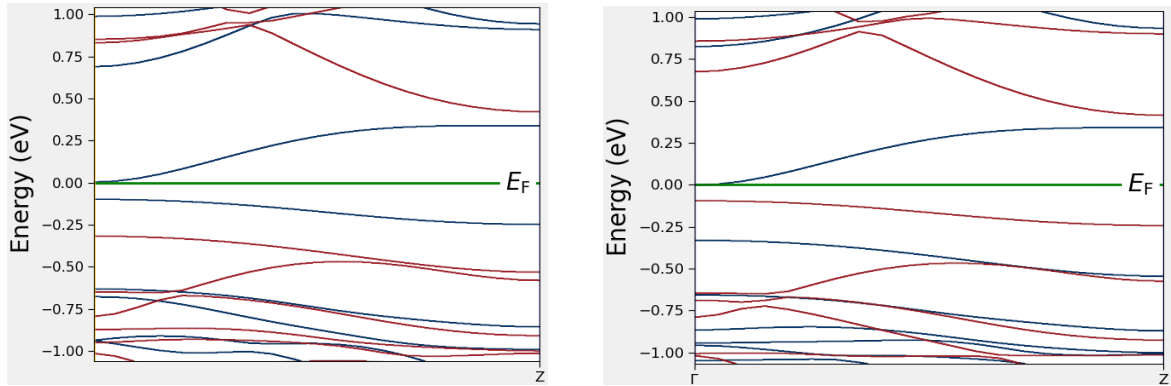

While not displaying spin gaps that may give rise to current spin polarization, the cases of N or O atom adsorption exhibit further modifications of the bands closest to the Fermi energy, confirming that the addition of adsorbate atoms significantly modifies the low-energy band structure of zTMDCNRs. It should be noted that for 5-zWS$_2$NR+O, the two frontier bands cross the Fermi energy, giving rise to a metallic ribbon.

*Figure 5:* Adsorption equilibrium geometries of 5-zWS2NR+*ad,* where panels (a) – (e) refer to *ad* = H, B, C, N, O, respectively. Also shown are isovalue representations of the corresponding spin density distribution. The isosurface parameter is 0.05 1/Å$^3$.

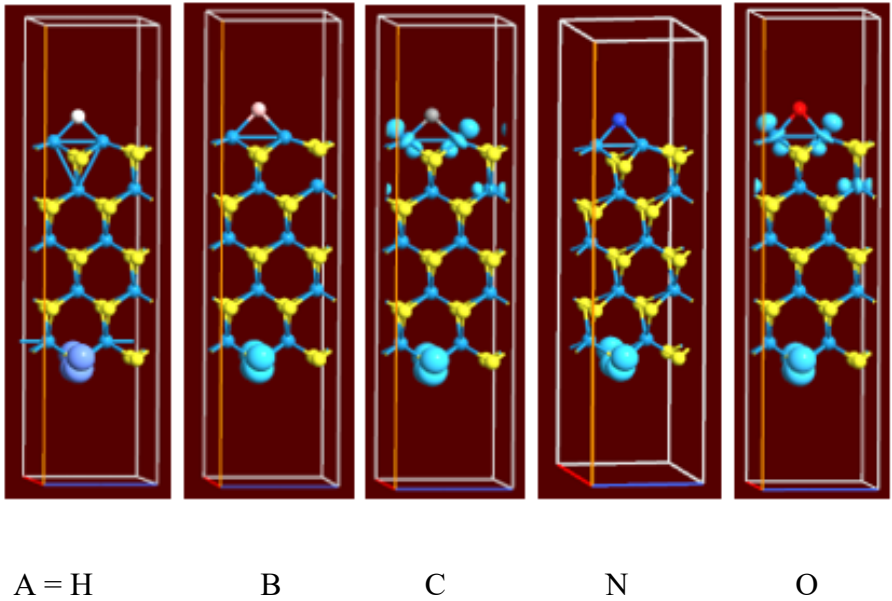

| A = H | B | C | N | O |

The Janus structures 5-zWSeSNR+H and 5-zWSeSNR+C (Figure 3(i) and (j), respectively) share salient features with their 5-zWS$_2$NR counterparts (Figure 3(c) and (e)). For 5-zWSeSNR+H (5-zWS$_2$NR+H), a spin gap opens in the conduction regime adjacent to $E_F$. For 5-zWSeSNR+C (5-zWS$_2$NR+C), the highest valence and the lowest conduction band define a spin gap that extends through an energy interval of about 0.5 eV.

Various signature features of 5-zWXYNR+n$ad$ (X, Y = S, Se; n = 0,1; $ad$ = H, B, C, N, O) band structure profile may be clarified in terms of the Bloch functions associated with the frontier orbitals. Thus, for $ad$ = B, C, N, O, the HVB turns out to depend only weakly on the adsorbate species, being composed mostly of S(3p) contributions located at the bottom edge. While containing, in all cases considered, a substantial W component, the LCB is found to vary significantly with the adsorbate which exerts a sensitive impact on the band structure close to $E_F$, determining both the energy gaps and the spin gaps of the ribbons. The Bloch function analysis of the frontier bands is described in greater detail in section *S-B* of the Supporting Information file.

(ii) Transmission spectra of 5-zWXYNR+*ad* (X, Y = S, Se; *ad* = H, C, N, O)

Building on our results on the band structure and the related electronic states, in the following step we compare the various 5-zWXYNR+*ad* (X, Y = S, Se; *ad* = H, C, N, O) systems in terms of their transmission spectra and emphasize that the band structure acts here as a useful predictor of electrode spin transmission features. Figure 6 displays the transmission spectra for some of the electrode systems discussed in the context of Figure 3. All transmission data were obtained by use of the HSE06 functional.

In accordance with the band structure of 5-zWS$_2$NR, we find a pronounced transmission gap of $\Delta E \sim 0.25$ eV around $E = E_F$ for this system, limited by a spin alpha LVB and a near-degenerate pair of two bands with opposite spin assignments in the valence regime (compare Figure 6(a) with Figure 3(b)). Figure 6(b) refers to 5-zWS2NR+H. Here, the narrow spin gap in the conduction region of 5-zWS2NR+H is clearly reflected by a zone where the transmission is confined to alpha spin orientation, i.e., T (alpha, E) = 1, T (beta, E) = 0 for $0.25$ eV $\leq E \leq 0.50$ eV. In the valence region, however, we find the same transmission onset energy for both spin orientations, resulting in zero net spin transmission for $E \leq E_F$. For 5-WS2NR+B (Figure 6(c)), one finds a sizable transmission gap centered on $E_F$, surrounded by near-coinciding spin alpha and beta transmission maxima, indicative of vanishing net spin polarization. These features are in keeping with the respective band structure (Figure 3 (d)). A C adatom gives rise to two spin beta transmission channels that derive from the highest valence and the lowest conduction band for 5-zWS2NR+C (compare with Figure 3(e)), both with spin beta orientation, and well separated from alpha bands.

Based on these findings, a high degree of spin polarization may be predicted for 5-WS2 with a C adatom. Very similar observations are made for the Janus structure 5-zWSeSNR+C (Figures 3(j), 6(h)).

*Figure 6:* Transmission spectra of 5-zWXYNR (X, Y = S, Se) electrode systems in the interval [-1.0, 1.0] eV. Shown are (a) 5-zWS$_2$NR (AFM), (b) 5-zWS$_2$NR+H (AFM), (c) 5-zWS$_2$NR + B (FM), (d) 5-zWS$_2$NR+C (FM), (e) 5-zWS$_2$NR+N (FM), (f) 5-zWS$_2$NR+O (FM), (g) 5-zWSeSNR+H (AFM), (h) 5-zWSeSNR+C (FM). The red (blue) line corresponds to spin alpha (beta) orientation.

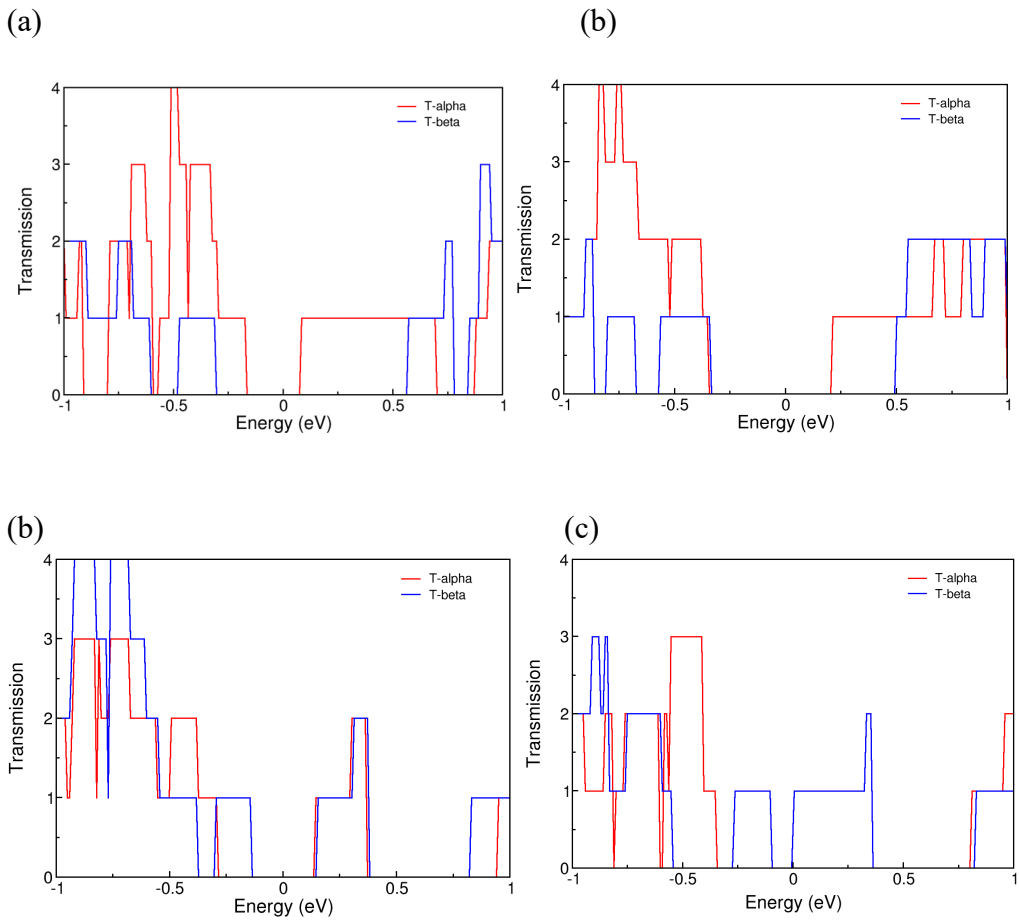

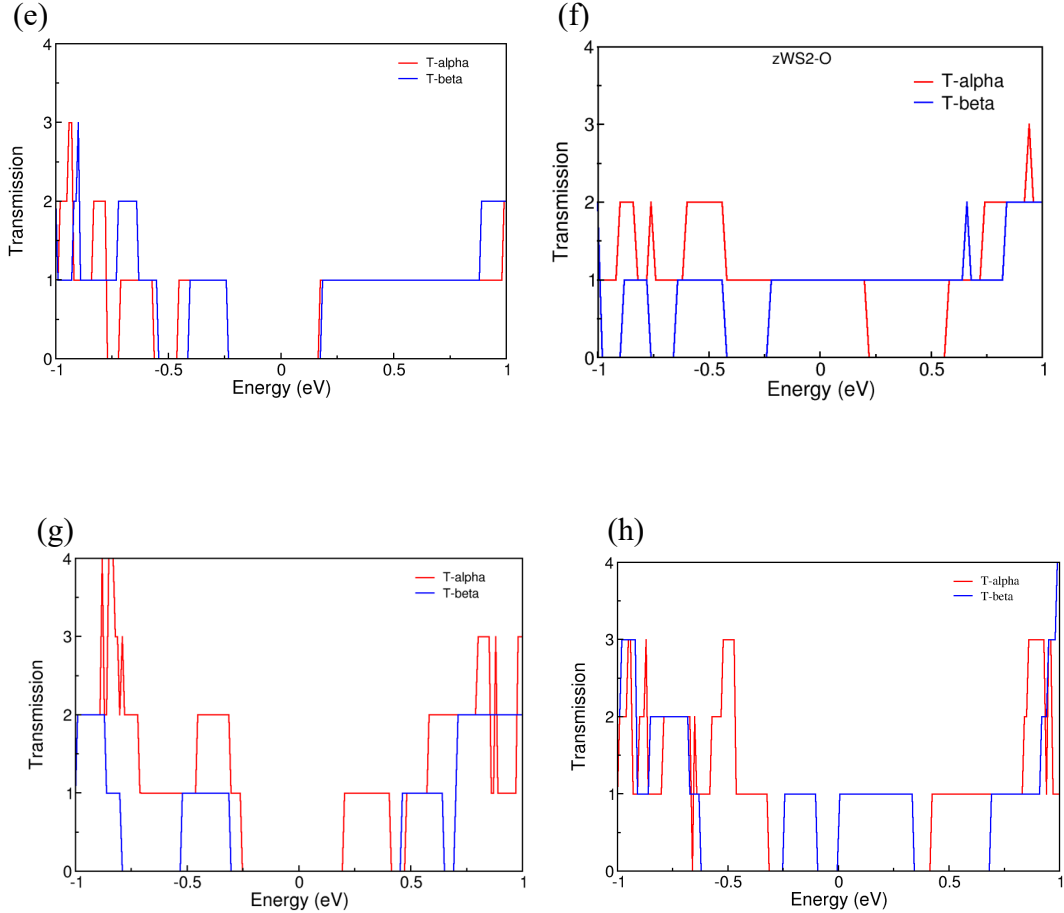

(B) Electrodes of the form 5-zWXYNR+2*ad* (X, Y = S, Se; *ad* = H, B, C,N,O)

Adding a second adatom to the top edge changes the electrode unit cell in terms of its geometric, electronic, and magnetic properties. With respect to the geometric structure of the 5-zWXYNR+2*ad* (X, Y = S, Se; *ad* = H, B, C, N, O) unit cell, we record a slight longitudinal expansion or contraction of the 5-zWXYNR+*ad* reference cell. The sign and size of this effect vary with the adatom element (see the respective entry in Table 3). In particular, cage expansion is found in the case of *ad* = B, C. Here, a staggered edge adsorption pattern emerges, associated with alternatingly extended and compressed distances between adjacent W centers in the top edge of the ribbon, and a correlated variation of the adatom height. The systems 5-zWS$_2$NR+2C and 5-zWSeSNR+2C display the additional geometric feature that the lower one of the two adatoms bonds to a triangular W-S-W substructure at the top edge, rather than to the two top W atoms alone (see Figure 7, (c))

*Figure* 7: Adsorption equilibrium geometries of zWS$_2$NR+2*ad*, with *ad* = (a) H, (b) B, (c) C, (d) N, (e) O. Left panels: sideview of the unit cell, right panel: view along the periodic coordinate. Also shown are isovalue representations of the respective spin density distributions. All systems are shown in their FM phases. The isovalue is 0.05 1/Å$^3$.

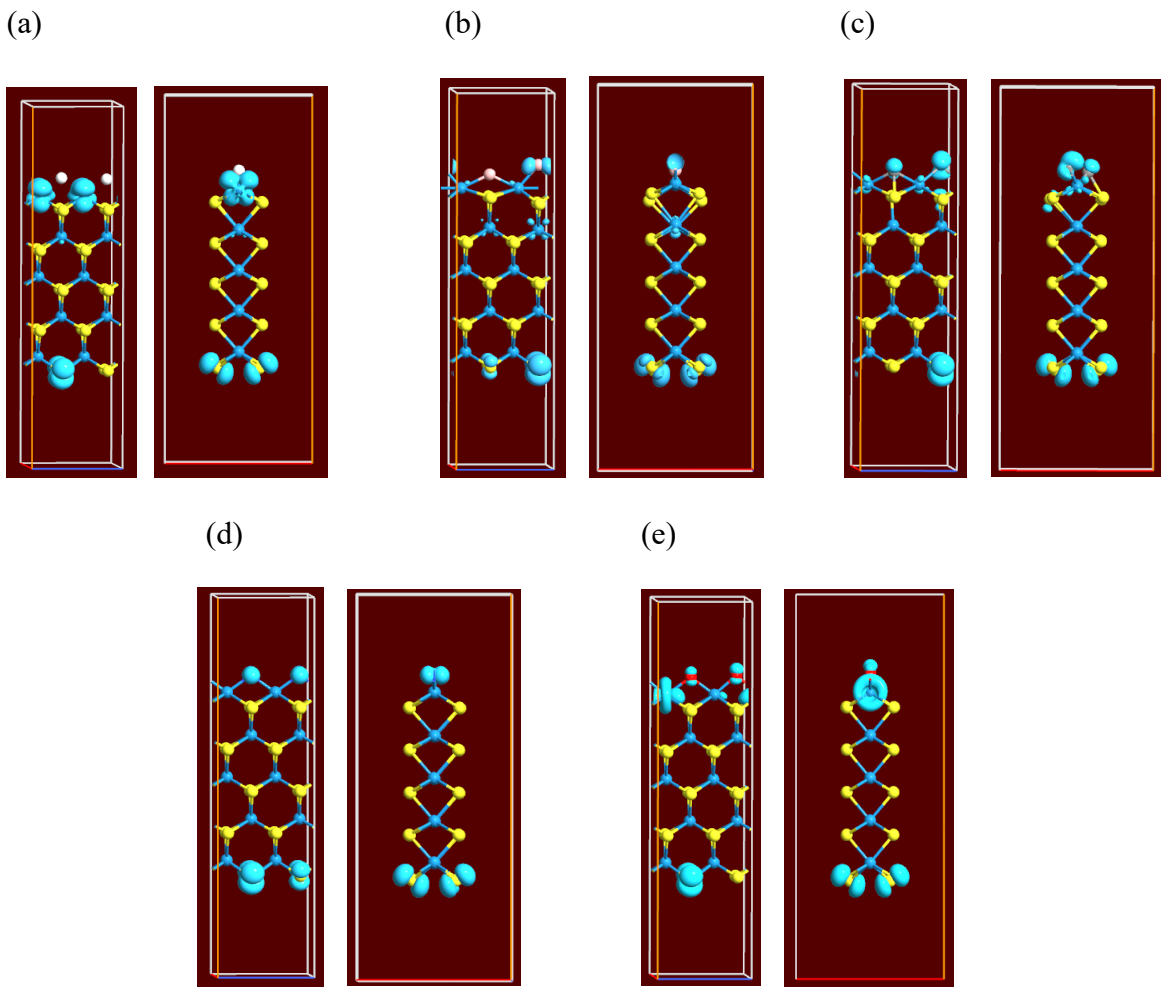

Various features observed for single-atom adsorption carry over to the double-atom adsorption case. Thus, the equilibrium energy difference between FM and the AFM configurations is, with the exceptions of 5-zWS$_2$NR+2H and 5-zWS$_2$NR+2C, once again in, or below, the 1 meV regime. Likewise, once more, we identify systems that display major rearrangement of the band structure in the E < E$_F$ regime when undergoing transitions between FM and AFM order, while their respective ground states are near-degenerate. This behavior is found for A = B, N (see Figure 9). For A = B, C, O, FM order turns out to be preferred by a small energy margin, while the AFM phase is more stable for A = H. Degeneracy of the two magnetic phases within ΔE = 0.1 meV is obtained for A = N. Figure 8 shows the band structure profiles of 5-zWXYNR+2*ad* (X, Y = S, Se, A = H, B, C, N, O) in their more stable magnetic phase, Figure 9 contrasts AFM and FM band structures for three selected systems (5-zWS2+2*ad* with A = B, N; 5-zWSeSNR+2C).

From Figure 8 and Table 3, one also notes marked differences between the two types of edge adsorption. In particular, the band structure profiles of 5-zWXYNR+2*ad* (X, Y = S, Se; A = H, B, C, N, O) show that adding a second adatom turns several of the single-adsorption systems metallic, as found for the cases of *ad* = B, N. Furthermore, in contrast to single-atom adsorption (see Table 1), and excepting the case *ad* = H, the two adatoms of 5-zWXYNR+2*ad* exhibit substantial degrees of spin polarization, as indicated by the net spin population values listed in Table 3. For *ad* = B, C, the geometric asymmetry between the two adatoms corresponds to a distinct spin polarization difference. We note that the largest degrees of spin polarization are found for 5-WS2NR+2C and 5-WSeSNR+2C, and are associated with the adatom that attaches to a W – S – W motif.

When comparing Figure 8 with Figure 2, we notice that the addition of a second adsorbate atom induces substantial changes in the band structure profiles in some cases (A = B, N, O), while the overall pattern of the bands close to E = E$_F$ is largely preserved in others (A = H, C). The latter group may be exemplified by 5-zWS$_2$NR+2C. The HVB↑/↓ consists here of S(3p) bottom edge states. In contrast to the case of single atom adsorption, the HVB↑ and HVB↓ bands are near-degenerate, eliminating the spin gap for $E < E_F$, as found for 5-zWS$_2$NR+C. The LCB is composed of top edge states that mix W(5d), S(3p), and C(2p) components, with contributions from both C atoms. The respective Bloch function at k = 0 is shown in Figure 10 (b) and compared with the LCB Bloch function for the single-atom adsorption system. The essential change from single to double adsorption is the addition of an electronic population at the site of the added C atom. Comparison between the LCB states of the units with one or two B atoms reveals a more extensive rearrangement (see Figure 10 (a)). For the latter, the electronic population at the top edge is largely carried by the added (i.e., the higher) B atom, while the former is characterized by a bonded configuration that includes all atoms at the ribbon top.

*Figure 8:* Band structures of 2-cell 5-zWXYNR+2ad (X, Y = S, Se, ad = H, B, C, N, O) electrode systems in the energy range [-1.0, 1.0] eV. Shown are: (a) 5-zWS$_2$NR+2H, (b) 5-zWS$_2$NR+2B, (c) 5-zWS$_2$NR+2C, (d) 5-zWS$_2$NR+2N, (e) 5-zWS$_2$NR+2O, (f) 5-zWSeS+2H, (g) 5-zWSeS+2C. In each case, the magnetic configuration of lowest energy is selected. All structures were computed using the HSE06 functional. Bands with spin alpha (beta) polarization are shown in red (blue).

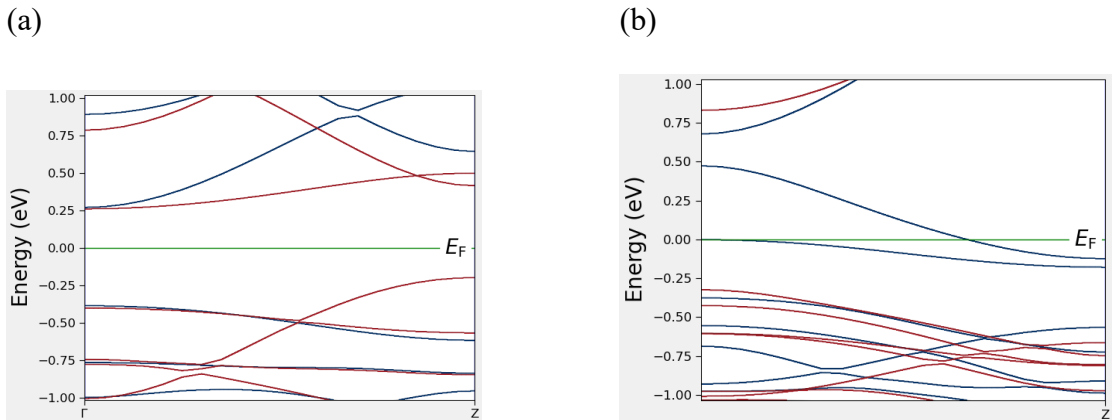

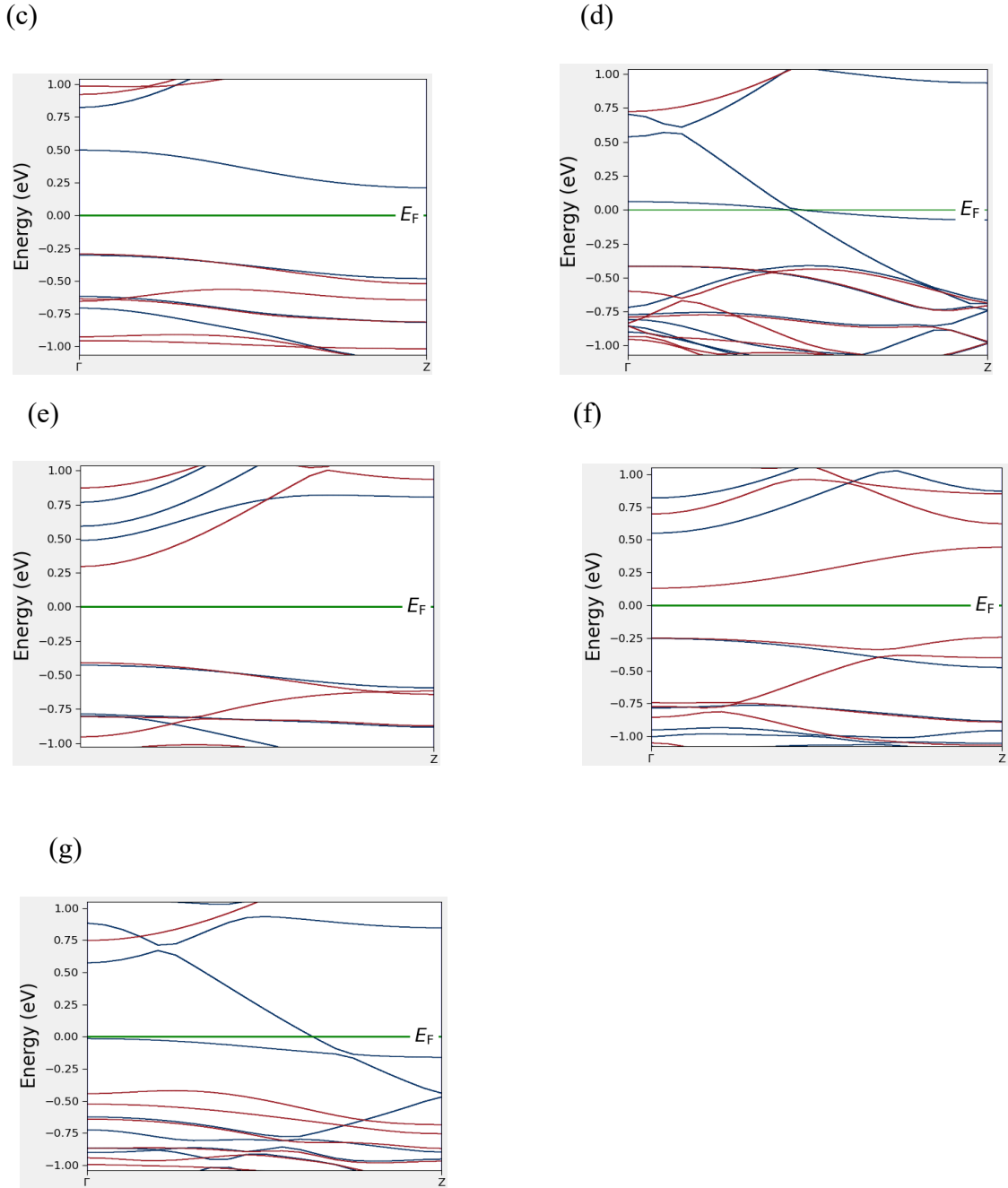

*Figure 9:* Band structures of 5-zWS2NR+2*ad* (*ad* = H, B, N) electrode systems in the energy range [-1.0, 1.0] eV. The left (right) panels show the band structure profiles under FM (AFM) conditions. Included are 5-zWS2+2H (a), 5-zWS2+2B (b), and 5-zWS2+2N (c).

(a)

FM            AFM

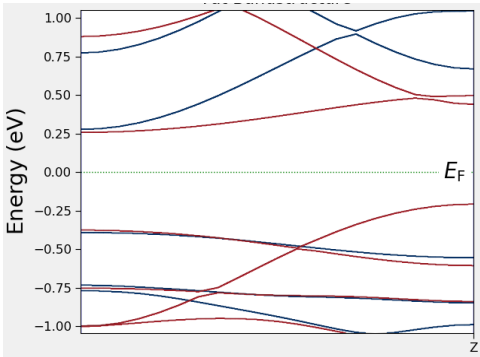 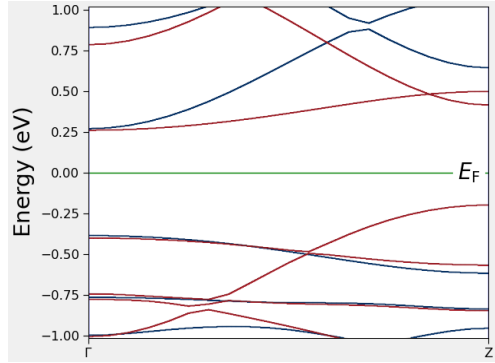

(b)

FM            AFM

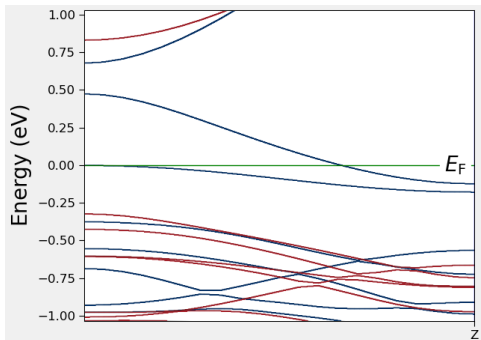 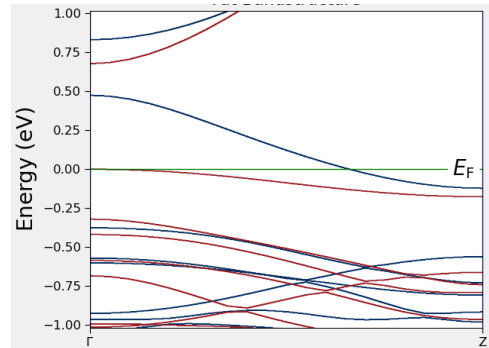

(c)

FM            AFM

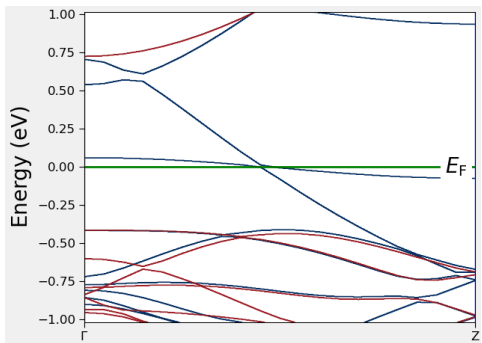 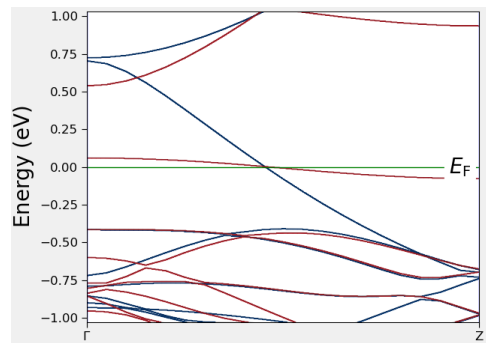

*Figure* 10. Comparison between the LCB Bloch functions of 5-zWS$_2$NR+*ad* (left panels) and 5-zWS$_2$NR+2*ad* (right panels) at k = 0. (a) *ad* = B, (b) *ad* = C.

(a)

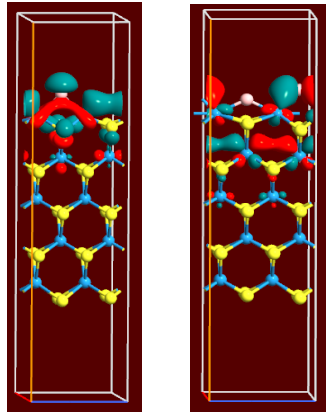

(b)

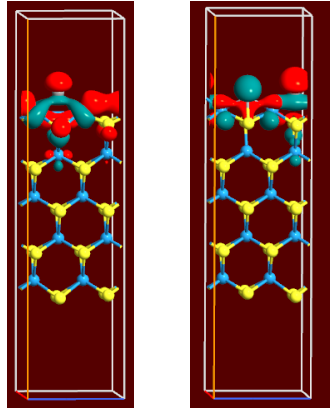

*Table 3*: Geometric, energetic, and magnetic parameters of double-cell 5-zWXYNR+2*ad* (X, Y = S, Se; A = H, B,C,N,O)  2-cell 5-zWXYNR (X, Y = S, Se).

| System | | E(HSE) | Change of longitudinal lattice parameter [%] | Net spin population on A1, A2 |
|---|---|---|---|---|
| 5-zWS$_2$NR+2H | AFM | -26031.5812 | -0.96 | -0.01, -0.02 |
| | FM | -26031.5769 | | -0.01, -0.02 |
| 5-zWS$_2$NR+2B | AFM | -26154.2134 | 2.22 | 0.28, -0.04 |
| | FM | -26154.2132 | | 0.28, -0.04 |
| 5-zWS$_2$NR+2C | FM | -26318.2102 | 0.48 | 0.41, 0.15 |
| | AFM | -26318.2076 | | 0.41, 0.15 |
| 5-zWS$_2$NR+2N | FM | -26556.9895 | -0.16 | 0.31, 0.31 |
| | AFM | -26556.9885 | | 0.31, 0.31 |
| 5-zWS$_2$NR+2O | FM | -26894.1244 | -0.96 | 0.07, 0.12 |
| | AFM | -26894.1244 | | |
| 5-zWSeSNR+2H | AFM | -55154.9182 | -1.70 | -0.01, -0.02 |
| | FM | -55154.9172 | | -0.01, -0.02 |
| 5-zWSeSNR+2C | FM | -55442.0563 | +0.78 | 0.35, 0.32 |
| | AFM | -55442.0562 | | 0.35, 0.32 |

## 3.2 Devices

Spin-resolved current computations were performed on device structures based on the 5-zWXYNR+n*ad* (X, Y = S, Se; n = 0, 1, 2; A = H, B, C, N, O) electrode units discussed above. In each case, the equilibrium geometry was obtained for the relevant structural prototype (see Figure 2(e, f)) by SGGA-PBE optimization. This was followed by a single-point current calculation on the SGGA-HSE level, yielding the current components with spin-alpha (spin-beta) polarization, I↑(I↓). In all cases, results were recorded for both the FM and the AFM alternative imposed on the initial state.

As a reference, we present in Figure 11(a, b) the current–voltage profiles for the pure 5-zWS2NR ribbon in the voltage interval [0.00, 18.0] mV for both spin orientations. The comparison between alpha- and beta-spin current demonstrates that the former prevails over the latter by more than three orders of magnitude, reflecting a distinct current spin polarization effect. We point out that the alpha component displays a negative differential resistance (NDR) effect in the voltage region $6.0 \leq V \leq 7.5$ mV, while the beta component exhibits the expected rise of the current as a function of the voltage (Figure 8 a(b)) The NDR feature is related to the change of the longitudinal electric field as the voltage increases, specifically its impact on the transmission function. This effect is demonstrated in Figure 11(c). Shown are the alpha spin transmission functions close to the transmission edge in the $E > E_F$ regime (compare with Figure 7) at two bias values, 6.0 and 7.5 mV. This implies a widening of the 5-zWS$_2$NR transmission gap for the latter versus the former voltage, which results in the non-Ohmic behavior of dropping rather than rising current upon voltage increase. For beta transmission, no such behavior is found, in accordance with the alpha character of the 5-zWS2NR frontier bands (see Figure 3(b)). Thus, 5-zWS$_2$NR gives an example for a spin selective NDR effect. In the context nanoelectronic circuits, NDR behavior is relevant primarily as it allows for high-speed switching [49], associated with fast on/off transitions, as small voltage changes can cause large changes in current. Thus, spin-selective NDR enables spin-resolved current control, which is essential for spin diodes and transistors, i.e., devices that respond differently to spin-up vs. spin-down electrons.

Similar arguments apply to 5-zWSeSNR, with the difference that here, the NDR effect is observed in the beta instead of the alpha moiety (Figure 11, (d) – (f)). It is not surprising that the NDR responses in the two compared systems deviate from each other. In contrast to 5-zWS$_2$NR, where the longitudinal field along the transmission element governs voltage-dependent deformations of the transmission spectrum, 5-zWSeSNR also exhibits a vertical gate field, induced by the difference between the two chalcogen groups and the associated electronegativity gradient in the direction perpendicular to the W layer.

*Figure 11*: NDR effect for the alpha current through 5-zWS$_2$NR and 5-zWSeSNR. (a) 5-zWS$_2$NR alpha current voltage profile. The current decreases as the voltage increases from 6.0 to 7.5 mV; (b) 5-zWS$_2$NR beta current voltage profile, displaying regular behavior; (c) transmission values close to the E > E$_F$ edge of the alpha transmission function at V = 6.0 mV and V = 7.5 mV; (d) 5-zWSeSNR alpha current voltage profile, displaying regular behavior; (e) 5-zWS$_2$NR beta current voltage profile. The current decreases as the voltage increases from 10.0 to 12.5 mV; (f) transmission values close to the E > E$_F$ edge of the alpha transmission function at V = 6.0 mV and V = 7.5 mV.

(a)

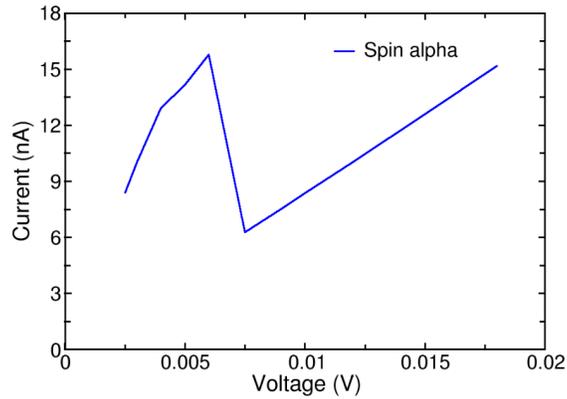

(b)

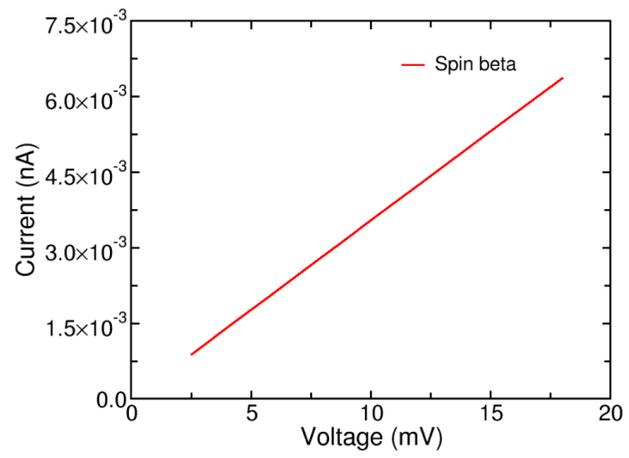

(c)

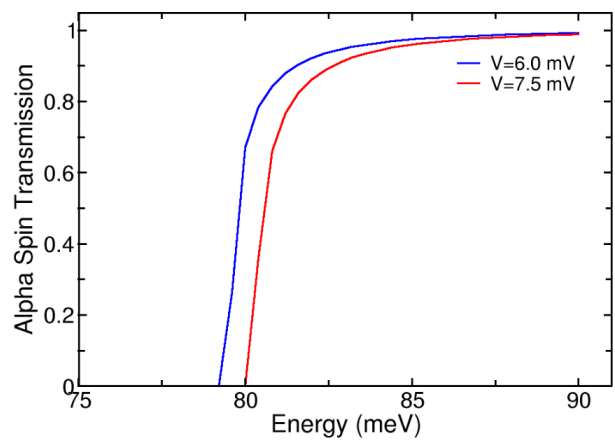

(d)

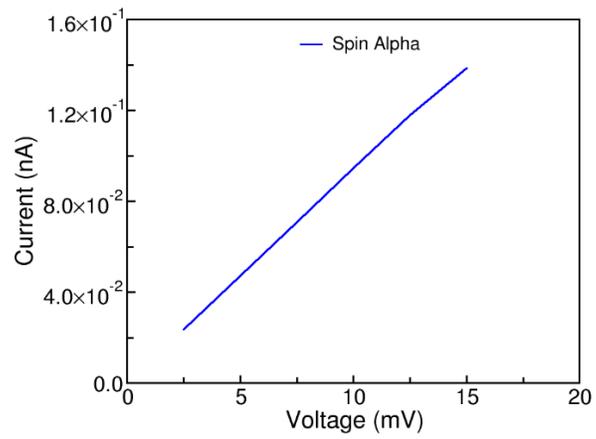

(e)

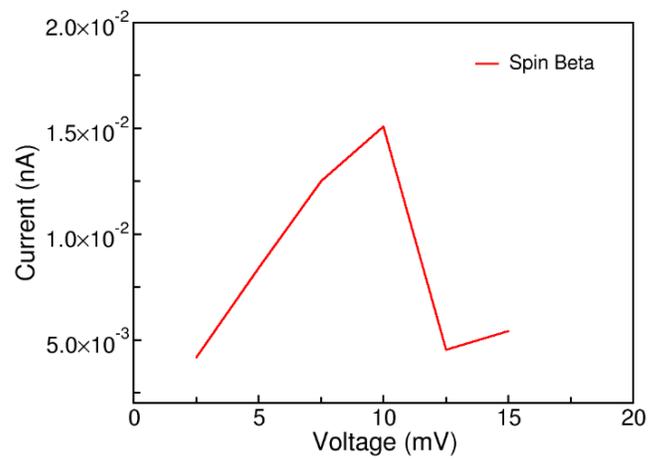

(f)

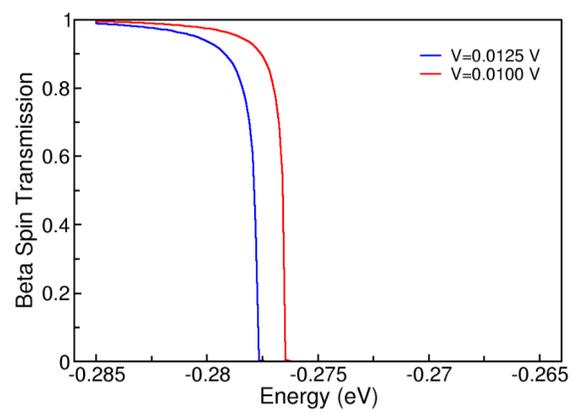

For quantitative assessment of the spin-filter property, we evaluate the magnetocurrent ratio, MCR:

$$\text{MCR} = [I(\uparrow) - I(\downarrow)]/[I(\uparrow) + I(\downarrow)] \qquad (5)$$

Table 4 lists the MCR values for devices of the type 5-zWXYNR+n$ad$ (X, Y = S, Se; n = 0, 1; $ad$ = H, B, C, N, O), along with the total current, $I_{tot} = I_\uparrow + I_\downarrow$, through the respective device at two selected voltages across the transmission element, V = 5.00, 10.0 mV. From test calculations, the device length of six-unit cells (see Figure 2, e, f) turned out to be sufficient to yield the asymptotic MCR (see section *S-D* of the Supporting Information file).

As documented by the entries for X, Y = S, n = 0, 1, and $ad$ = H, the pure (the H decorated) 5-zWXYNR device gives rise to complete (close to complete) current spin polarization, in agreement with the transmission behavior of the corresponding electrode units (Figure 6 a, b). In both cases, the LCB is an alpha band, whereas the HVB↑ and the HVB↓ are near-degenerate (Figure 3b, c). The high and positive MCRs for both systems are strongly determined by the LCB, reflecting the fact that E(LCB) – $E_F$ << $E_F$ – E(HCB). Despite the strong spin polarization effect, the potential practical use of the pure and the H-decorated 5-WS$_2$NR devices is severely limited. Thus, the high reactivity of the unprotected metallic top edge makes it challenging to use pure 5-WS$_2$NR in the laboratory. 5-WS$_2$NR+H devices, on the other hand, are unfeasible as elements of spintronics networks due to their small current magnitudes in the low voltage regime, a direct consequence of their sizeable electronic gaps around $E_F$. The conclusions drawn for 5-WS2NR+H carry over to its Janus counterpart, 5-WSeSNR+H.

Distinctly larger current is found for 5-zWXYNR+C, with $I_{tot}$ in the order of 1 nA. As in the cases discussed above, the MCR is here determined by the thermally weighted average of the FM and the AFM solution, where, on account of their small difference in total energy, both alternatives contribute significantly. From Table 2 (f), the respective band structures show markedly different profiles, the HVB and LCB displaying equal (opposite) spin orientations for the FM (AFM) component. The MCR values for the competing structures, however, turn out to be very close to each other, with a thermal average of -98.9% (-97.9%) for V = 5.00 mV (10.0 mV). As for the pure 5-WS2NR and 5-WS2NR+H, the almost perfect current spin polarization is due to the asymmetric placement of the two frontier bands, the LCB nearly coinciding with $E_F$ in the small $k$ regime, while the minimum separation between HVB is close to 0.1 eV. Analogous observations are made for the corresponding Janus structure, 5-zWSeSNR+C.

For A = B, N, and O, near-vanishing MCRs are obtained. Again, this finding is rationalized by analyzing the respective band structure profiles in terms of spin gaps. As the ribbons with A = B, N do not exhibit any spin gaps, the MCR reduces to zero in these cases. Likewise, the symmetric arrangement of an alpha (E > $E_F$) and a beta (E < F) spin gap leads to the expectation that no current spin polarization will arise for this system in the low-voltage range, as confirmed by the MCR result.

We investigated the spin-filtering capacity of 5-zWXYNR+2*ad* (X, Y = S, Se; A = H, B, C, N, O) for selected systems (*ad* = H, C). As in the single-adsorbate case, evaluating the current spin polarization of the double-adsorbate systems needs to include both the FM and AFM configurations, wherever their energy difference is in the meV regime, which holds for all.

The transmission features of 5-zWS$_2$NR+2H and 5-zWSeSNR+2H (AFM) are closely related to those of the ribbons with a single H adsorbate. However, in contrast to 5-zWS$_2$NR+H, the spin alpha LCB of 5-WS2NR+2H coincides for small k with the second-lowest conduction band, which carries the opposite spin orientation (see Figure 8(a)). This ambiguity deteriorates the MCR from 100% to 73.5 %. For 5-zWSeSNR+2H, a clear separation between the lowest and second-lowest conduction band is found. The respective MCR is maximum and thus close to the single-adsorption case.

Table 4: Total currents and Magnetic Current Ratios (MCRs) for 5-zWXYNR+*ad* (X, Y = S, Se; *ad* = H, C) devices at V = 5.0, 10 mV. The results refer to the more stable of the two compared magnetic configurations, FM and AFM.

| *System* | *Voltage(mV)* | $I_{tot} = I_\uparrow + I_\downarrow$ *(nA)* | *MCR (%)* |
|---|---|---|---|
| 5-zWS$_2$NR | 5.00 | 14.2 | 100 |
| | 10.0 | 8.35 | 100 |
| 5-zWS$_2$NR+H | 5.00 | 3.80 10$^{-2}$ | 98.5 |
| | 10.0 | 7.63 10$^{-2}$ | 98.3 |
| 5-zWS$_2$NR+C | 5.00 | 1.73 | -100 |
| | 10.0 | 3.33 | -100 |
| 5-zWSeSNR | 5.00 | 5.57 10$^{-2}$ | 70.2 |
| | 10.0 | 11.1 10$^{-2}$ | 73.9 |
| 5-zWSeSNR+H | 5.00 | 3.80 10$^{-2}$ | 98.5 |
| | 10.0 | 7.63 10$^{-2}$ | 98.3 |
| 5-zWSeSNR+C | 5.00 | 74.9 | -100 |
| | 10.0 | 130 | -100 |

Table 5: Total currents and Magnetic Current Ratios (MCRs) for 5-zWXYNR+2*ad* (X, Y = S, Se; ad = H, C) devices at V = 5.0, 10 mV. The MCR results for $I_{tot}$ and MCR are thermal averages between the two magnetic configurations, FM and AFM (T = 300 K)

| System | Voltage(mV) | $I_{tot} = I_\uparrow + I_\downarrow$ (nA) | MCR (%) |
|---|---|---|---|
| 5-zWS2NR+2H | 5.00 | 5.31 $10^{-2}$ | 73.2 |
|  | 10.0 | 10.6 $10^{-2}$ | 74.2 |
| 5-zWS2NR+2C | 5.00 | 3.86 $10^{-2}$ | -91.8 |
|  | 10.0 | 8.34 $10^{-2}$ | -91.8 |
| 5-zWSeSNR+2H | 5.00 | 3.68 $10^{-1}$ | 100 |
|  | 10.0 | 7.25 $10^{-1}$ | 100 |
| 5-zWSeSNR+2C | 5.00 | 2.69 $10^{2}$ | 76.0 |
|  | 10.0 | 5.07 $10^{2}$ | 76.1 |

## 4. CONCLUSIONS AND OUTLOOK

In this contribution, we propose a novel approach towards adapting zigzag nanoribbons based on transition metal dichalcogenides (5-zTMDCNRs) to spintronics applications, namely, functionalizing one of their zigzag edges with light, non-metallic adsorbate atoms. Specifically, optimizing the spin filtering capacity of zTMDC-based ribbons is achieved by leveraging atomic edge adsorption as a means to modify energy and spin gaps - zones in which the energy bands are dominated by a unique spin orientation - close to the Fermi energy.

As the electronic and spintronic transport properties depend sensitively on the band structure in the Fermi energy regime, an accurate representation of the electrode energy bands close to $E_F$ is crucial for this work. Thus, electronic structure calculations were performed by use of the hybrid

Heyd–Scuseria–Ernzerhof (HSE) functional. The non-equilibrium Green's function (NEGF) approach was applied to model spin and charge transport properties.

Here, 5-zWS$_2$NR and the Janus structure 5-zWSeSNR are chosen as TMDC reference systems. While the pure 5-zWS$_2$NR unit exhibits ground-state magnetism, it also displays an electronic gap of width 0.25 eV, compromising its charge and spin transport properties in the millivolt regime. In addition, the Bloch states of the weakly bound zTMDCNR bands are largely composed of incompletely saturated d-orbitals of the TM component and thus highly reactive, so that low chemical stability is expected for the pure zWXYNR species.

On the other hand, chemical edge modification by the addition of atomic adsorbates may stabilize the pure nanoribbons and, at the same time, restructure the low-energy regime of the 5-zWXYNR (X, Y = S, Se) energy band profile to yield more favorable transmission features.

Promising candidate units for use in spintronics were identified by inspecting the band structures and the associated spin-resolved transmission spectra of 5-zWXYNR electrode cells with light, non-metallic adsorbate atoms *ad*, where *ad* = H, B, C, N, and O. The main conclusions related to the (a) electronic and (b) magnetic reconstruction due to adsorption are the following:

(a) The 5-zWXYNR energy gap varies substantially with the type and the number of atomic adsorbates. Thus, attaching H, B, or N adsorbates widens the gap from its value for the pristine species (0.25 eV), but the gap narrows to 0.1 eV and even to zero for single adsorption of C or O atoms, respectively. It is found to vanish for double adsorption of N atoms.

(c) The spin at the adsorbing edge tends to be quenched for units of the type 5-zTMXYNR+*ad*, so that the magnetism of the ribbon is mostly localized at the opposite edge. For zTMXYNR+2*ad*, in contrast, both ribbon edges turn out to be magnetically active in several cases.

For all structures of potential interest for spintronics, the current-voltage profiles were computed, yielding direct information about the degree of current spin polarization in the considered voltage interval. The FM and AFM phases of all structures investigated are of almost equal stability, being separated by small energy margins in the meV range. This implies that the overall spin current polarization, as documented by the MCR, has to be computed as the thermal average of the two alternative magnetic states. From research presented here, two of the considered adsorption systems qualify as elements of nanospintronics by all three criteria – reduction of the 5-zWXYNR electronic gap, emergence of an extended spin gap in the environment of $E_F$, similarly high MCR values for FM and AFM order, namely 5-zWXYNR and 5-zWSeSNR with single adsorption of C atoms. For these cases, MCRs of maximum magnitude, reaching the limit of half-metallicity, are found consistently. We point out that a consequence of these results is a giant magnetoresistance effect for the transmission elements with *ad* = C. Furthermore, a spin-selective NDR effect was established for the pristine 5-zWXYNR and 5-zWSeSNR ribbons involving a marked current reduction upon a small voltage in the spin alpha (spin beta) system of the former (latter) unit.

In continuation of this work, we will investigate the impact of spin-orbit interactions, which has been shown to affect various TMDC structures of recent interest (see, e.g., [50]). Although the treatment of this effect is presently too demanding to be included here, it will be interesting to assess it in the future. Thus, the action of spin-orbit effects on the frontier bands of the systems considered here is of potential relevance for their current spin polarization features. This holds especially for bands that contain W atoms (i.e., mostly the LCBs which might be shifted or, in cases of coinciding bands with opposite spin orientations, split under the influence of spin-orbit interactions). The impact on S-atom dominated bands (mostly HVBs), in contrast, should be comparatively small.

Future extensions of this research will also include studies on the validity of the results presented here when disorder, both in the adsorbate subsystem and the 5-zWXYNR substrate, is taken into account. In addition, comparison will be made with experimentally accessible spin current carrier systems, such as graphene nanoribbons and their derivatives.